\documentclass[prl,twocolumn,showpacs,superscriptaddress]{revtex4}
\usepackage{amsmath}
\usepackage{graphicx}
\usepackage{dcolumn}
\usepackage{bm}

\begin{document}

\title{Precise measurements of radio-frequency magnetic susceptibility in (anti)ferromagnetic materials}
\author{M. D. Vannette}
\affiliation{Ames Laboratory and Department of Physics \& Astronomy, Iowa State University, Ames, IA 50011}
\author{A. Safa-Sefat}
\affiliation{Ames Laboratory and Department of Physics \& Astronomy, Iowa State University, Ames, IA 50011}
\author{S. Jia}
\affiliation{Ames Laboratory and Department of Physics \& Astronomy, Iowa State University, Ames, IA 50011}
\author{S. A. Law}
\affiliation{Ames Laboratory and Department of Physics \& Astronomy, Iowa State University, Ames, IA 50011}
\author{G. Lapertot}
\affiliation{D´epartement de Recherche Fondamentale sur la Mati`ere Condens´ee, SPSMS, CEA Grenoble, 38054 Grenoble, France}
\author{S. L. Bud'ko}
\affiliation{Ames Laboratory and Department of Physics \& Astronomy, Iowa State University, Ames, IA 50011}
\author{P. C. Canfield}
\affiliation{Ames Laboratory and Department of Physics \& Astronomy, Iowa State University, Ames, IA 50011}
\author{J. Schmalian}
\affiliation{Ames Laboratory and Department of Physics \& Astronomy, Iowa State University, Ames, IA 50011}
\author{R. Prozorov}
\affiliation{Ames Laboratory and Department of Physics \& Astronomy, Iowa State University, Ames, IA 50011}
\email[corresponding author:]{prozorov@ameslab.gov}

\date{11 May 2007}

\begin{abstract}
Dynamic magnetic susceptibility, $\chi$, was studied in several intermetallic materials exhibiting ferromagnetic, antiferromagnetic
and metamagnetic transitions. Precise measurements by using a 14 MHz tunnel diode oscillator (TDO) allow detailed insight into the field
and temperature dependence of $\chi$. In particular, local moment ferromagnets show a sharp peak in $\chi(T)$ near the Curie
temperature, $T_c$. The peak amplitude decreases and shifts to higher temperatures with very small applied dc fields. Anisotropic
measurements of CeVSb$_3$ show that this peak is present provided the magnetic easy axis is aligned with the excitation field. In a
striking contrast, small moment, itinerant ferromagnets (i.e., ZrZn$_2$) show a broad maximum in $\chi(T)$ that responds
differently to applied field. We believe that TDO measurements provide a very sensitive way to distinguish between local and
itinerant moment magnetic orders. Local moment antiferromagnets do not show a peak at the N\'eel temperature, $T_N$, but only a sharp
decrease of $\chi$ below $T_N$ due to the loss of spin-disorder scattering changing the penetration depth of the ac excitation
field. Furthermore, we show that the TDO is capable of detecting changes in spin order as well as metamagnetic transitions. Finally,
critical scaling of $\chi(T,H)$ in the vicinity of $T_C$ is discussed in CeVSb$_3$ and CeAgSb$_2$.
\end{abstract}

\maketitle


\section{Introduction}

Dynamic magnetic susceptibility, $\chi=\partial M /\partial H$, is an
important quantity to measure in magnetic materials. It couples directly to
the spin structure in the ordered state, critical fluctuations near the
transition, and the magnetic polarizability of the conduction electrons. The
sensitivity of techniques for measuring dynamic susceptibility varies
significantly with frequency and with the way the signal is obtained.
Amplitude - domain measurements with lock-in amplifiers operate at
relatively low frequencies and are prone to waveform distortions and phase
drift. Frequency - domain measurements are much more sensitive, but require
high-Q circuits and frequency scanning to find the resonance frequency. The
most sensitive probes utilize microwave cavity perturbation at typical
frequencies of $f\geq 1-100 \mathrm{\ GHz}$. However, at these frequencies,
nonlinear magnetic effects as well as the anomalous skin effect may be of
importance and interpretation of the results becomes difficult.

Tunnel diode oscillators (TDO) are well suited to fill the spectral gap
between these two extremes since they can be tuned to operate in the
radio-frequency (rf) range, as was suggested by Clover and Wolf \cite%
{clover-1970}. A TDO is essentially a self-resonating $LC$ tank
circuit biased by a tunnel diode. A properly constructed circuit
allows one to measure changes in the magnetic moment on the order of
a pico-emu. In addition, the excitation field is very small, of the
order of 20 mOe, which is very useful in cases of highly nonlinear
and/or hysteretic samples. These facts taken together suggest using
this instrument to measure magnetic susceptibility in the vicinity
of magnetic transitions where critical fluctuations are of
importance, but may be easily suppressed by fields of only a few
oersted (typical for conventional magnetometry). The major
development of the TDO as an instrument was reported by Van Degrift
in 1975 \cite{vandegrift-1975}, primarily to measure dielectric
constants of various materials. Earlier in that same year, Habbal,
\emph{et al}. described a simple TDO setup that could be used to
measure the rf susceptibility of samples over a broad temperature
range \cite{habbal-1975}. More recently, TDO's have been developed
to measure minute changes in the London penetration depth of small
superconducting samples \cite{prozorov-2006-2}. In addition, the
technique is sensitive enough to measure quantum oscillations in
resistivity via the normal skin depth \cite{prozorov-2006}.

Several groups have used TDO's to detect the ferromagnetic transition \cite
{fox-1986,parimi-2000,woods-2005}, however there has been little attention
paid to the vicinity of the transition itself. In this paper we demonstrate
that the TDO is a versatile and extremely sensitive quantitative probe of
ferromagnetic (FM), antiferromagnetic (AFM), and metamagnetic (MM)
transitions. This sensitivity proves particularly important when dealing
with questions related to small and very field-dependent features associated
with the critical fluctuations and delocalized magnetic moment of conduction
electrons.

\section{Experimental}

\subsection{Samples}

Single crystals of several intermetallic systems were used throughout this
study. All samples, except for ZrZn$_2$, were grown via high-temperature
solution growth \cite{canfield-1992,canfield-2001}. CeSb was grown out of
tin flux \cite{wiener-2000} while CeVSb$_3$ was grown out of excess Sb \cite
{sefat-2006}. [Ce, Sm]AgSb$_2$ were grown out of excess Sb and Ag \cite
{myers-1999,myers-1999-2}. GdFe$_2$Zn$_{20}$ was grown out of excess Zn \cite
{jia-2006}. ZrZn$_2$ was grown from a melt of ZrZn$_{2.006}$ to compensate
for zinc lost due to vaporization as described in Ref.%
\onlinecite{deReotier-2006}. Samples for TDO measurements were either cut
with a blade or used as grown and mounted on a 1 mm diameter by 15 mm long
sapphire rod. The TDO circuit and sample were enclosed in a vacuum can and
placed in the bore of a superconducting magnet with field range of 0-90 kOe.
The axis of the excitation coil and the static field were aligned. Both of
the RAgSb$_2$ samples were plates of dimensions $0.75 \times 0.75 \times 0.2$
mm with the c-axis perpendicular to the face. In these samples the c-axis
was also aligned with the coil axis. The CeSb sample was a rectangular slab
with dimensions of $0.25 \times 0.25 \times 0.5$ mm, the long axis aligned
with the [100] axis as well as with the coil axis. The Ce$_3$Al$_{11}$ and
CeVSb$_3$ samples were tapered needles with lengths of 1 mm. The aluminum
compound had a square base approximately 0.4 mm on a side, while the
triantimonide had a rectangular base approximately $0.4 \times 0.1$ mm. In
both of these samples the long axis coincided with the magnetic easy axis as
well as the coil axis. A small piece of the CeVSb$_3$ sample was cut from
the original sample in order to study the anisotropic response. This piece
had dimensions of $0.5 \times 0.2 \times 0.1$ mm with the 0.2 mm dimension
aligned with the c-axis as determined from Laue diffraction measurements.
The ZrZn$_2$ sample was roughly cubic with an edge length of 0.5 mm. The low
anisotropy of this compound eliminated the need to align any particular axis
with the coil. GdFe$_2$Zn$_{20}$ was a triangular prism. The base was
approximately $0.2 \times 0.2$ mm and the length was approximately 0.5 mm.
As with ZrZn$_2$, the low anisotropy of GdFe$_2$Zn$_{20}$ eliminated the
need for a particular crystal axis alignment. The long axis was aligned with
the coil. Samples used in magnetization and conventional resistivity
measurements where taken from the same growths as those used in the TDO
experiments. Magnetization measurements were performed in a \textit{Quantum
Design} MPMS operating down to 1.8 K with maximum applied field of 55 kOe.
Specific heat and conventional four-probe transport measurements were
performed in a Quantum Design PPMS system operating down to 1.8 K with a 90
kOe superconducting magnet.

\subsection{Principle of TDO operation}

The main component of a TDO is a tunnel diode. It has a heavily
overdoped, narrow ($\sim100$ $\mathring{A}$), p-n junction, so that
valence of p-side and conduction of n-side bands overlap. Above
certain bias voltage, forward bias results in a lesser overlap of
the two bands, thus reducing the tunneling current. As a result, the
device exhibits a bias region with negative differential resistance.
When biased to this voltage region, the diode can be used to drive
the tank circuit resulting in a self-resonating $LC$ oscillator. The
frequency is always at the resonance, which is the main advantage of
this device. Careful design and good thermal stability result in a
circuit that resonates in the rf range with a stability of
$0.05\mathrm{\ Hz}$ over hours to days
\cite{carrington-1999,prozorov-2000}. We report results obtained
with a TDO mounted in a $^3$He cryostat with a temperature range
from 0.3-150 K operating at a frequency of approximately
$14\mathrm{\ MHz}$. A dc external field up to 90 kOe can be applied
to study field-dependent
properties. A sample is placed in a coil which acts as the inductor in the $%
LC$ tank circuit. The shift of the resonant frequency is directly related to
the dynamic susceptibility, $\chi$, of the sample.

The frequency of oscillation for an inductor-capacitor circuit with
inductance $L$ and capacitance $C$ is given by
\begin{equation}
f_0={\frac{1}{{2\pi \sqrt{LC}}}}.
\end{equation}
If the inductance changes a small amount to $L+\Delta L$ the new frequency
can be written as
\begin{equation}
f_0+\Delta f={\frac{1}{{2 \pi \sqrt{(L+\Delta L)C}}}}.
\end{equation}
Expanding this expression for small values of $\Delta L$ gives, to first
order,
\begin{equation}
\Delta f\approx -{\frac{1}{{2}}}{\frac{\Delta L }{{L}}}f_0.
\end{equation}
If a magnetically active sample is responsible for the change in inductance,
one can rewrite the above expression as
\begin{equation}
{\frac{\Delta f }{{f_0}}} \approx -{\frac{1}{2}} {\frac{V_s}{{V_c}}}4 \pi
\chi_m,
\end{equation}
where $\chi_m$ is the measured susceptibility, $V_s$ is the sample volume,
and $V_c$ is the inductor (or coil) volume. In the above expression, $f_0$
corresponds to the empty coil resonant frequency and $\Delta f$ is the shift
in resonance caused by inserting the sample.

If there is non-zero demagnetization factor, $N$, measured susceptibility, $%
\chi_m$, is related to the true (zero-demagnetization) susceptibility, $%
\chi_t$, via

\begin{equation}
\chi_m={\frac{\chi_t }{{1+4 \pi N \chi_t}}}.
\end{equation}

Thus, the change in resonant frequency is directly proportional to the
dynamic magnetic susceptibility of any sample placed in the coil. An
increase in magnetic susceptibility results in a decrease in resonant
frequency and \emph{vice versa} for a diamagnetic response.

Measured susceptibility, $\chi_m$, is composed of two parts: (1) the
magnetic response of the bulk to an applied field and (2) the screening due
to the skin effect in metals. The two effects add to give the total $\chi$.
\begin{equation}
\chi_{total}=\chi_{mag} + \chi_{skin}.
\end{equation}
The usual interpretation of $\chi_{mag}$ comes from $M = \chi_{mag} H$. This
may be paramagnetic or diamagnetic. $\chi_{skin}$ is always diamagnetic. In
normal metals this is conventional skin depth, whereas in superconductors
this contribution is associated with London penetration depth. For
conducting ellipsoidal samples $\chi_{skin}$ can be written as
\begin{equation}  \label{diamag}
\chi_{skin}=-{\frac{1 }{{4 \pi}}} (1- {\frac{\delta }{{2R}}}\tanh{\frac{2R}{{%
\delta}}}).
\end{equation}
Here $R$ is the characteristic dimension of the sample perpendicular to the
axis of the coil and $\delta$ is the skin depth \cite{hardy-1993}. Skin
depth is related to resistivity, $\rho$ as below
\begin{equation}
\delta=\sqrt{\frac{{2 \rho}}{{\mu f}}},
\end{equation}
where $f$ is the oscillator frequency and $\mu$ is the magnetic permeability
of the material \cite{jackson}. If $R$ is much larger than $\delta$ then
equation \ref{diamag} may be written as
\begin{equation}
\chi_{skin}=-{\frac{1 }{{4 \pi}}}(1-{\frac{\delta }{{2R}}}).
\end{equation}

If necessary, conventional resistivity measurements can be made and the
effect of diamagnetic screening can be subtracted. While some of our samples
do show a change in resistivity, in the vicinity of the phase transition the
dominant contribution comes from the magnetic component. Further, over the
field range of interest in the vicinity of the phase transition (typically
less than 1 kOe) for the samples presented here magnetoresistance is
negligible. Therefore, we can also subtract a ``high" field data run if we
need to eliminate the skin-depth background.

\section{Results}

\subsection{Probing Magnetic Transitions}

The materials CeVSb$_3$, CeAgSb$_2$, ZrZn$_2$, and GdFe$_2$Zn$_{20}$ exhibit
ferromagnetic order at approximately 4.5, 9.8, 28, and 86 K respectively.
SmAgSb$_2$ is an antiferromagnet with a N\'eel temperature of approximately
10 K. CeVSb$_3$ and CeAgSb$_2$ are local moment magnets with a reported
saturated moments of 1.80 $\mu_B$ \cite{sefat-2006} and 0.4 $\mu_B$ \cite
{myers-1999} per cerium atom. ZrZn$_2$ is a weak itinerant ferromagnet which
is also highly unsaturated, showing a large relative increase in moment up
to fields of 5.5 T. Its reported ordered moment is about 0.17 $\mu_B$ per
formula unit \cite{yelland-2005}. The ground state of GdFe$_2$Zn$_{20}$ is
reported as a mixed Gd-local moment/Fe-itinerant system with an unusually
high $T_C$ attributed to a highly polarizable conduction electron
background. DC magnetization measurements show that it carries a moment of
about 6.5 $\mu_B$ per formula unit with an induced Fe moment opposite the Gd
\cite{jia-2006}. We present measured $\chi$ for each material is several low
fields.

\subsubsection{CeVSb$_3$}

The ternary rare earth compound CeVSb$_3$ undergoes a ferromagnetic
transition at 4.5 K with moments aligned along the crystallographic c-axis
\cite{sefat-2006}. The material grows as elongated plates and fractures into
needles with the long axis along the ferromagnetic easy axis. Figure \ref%
{cevsb3} shows the resonator frequency shift relative to the high
temperature value as temperature was increased from 3.5 to 7 K in several
applied magnetic fields. The zero field response is quite sharp and very
large. A field of 125 Oe is sufficient to suppress the susceptibility peak
by 70\%. A 1 kOe field almost completely suppresses the peak, and shifts it
to above 5 K. Similar behavior has been reported for an amorphous Fe-Ni-B-Si
alloy using a compensated coil ac susceptometer (\cite{drobac-1996} and \cite
{drobac-1998}) operating at 231 Hz, but the zero field data shown here has a
much sharper response.

\begin{figure}[htb]
\begin{center}
\includegraphics[width=9cm]{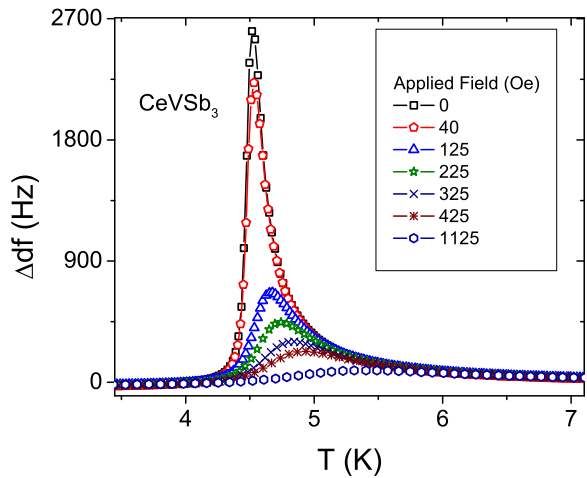}
\end{center}
\caption{(color online) Change in resonant frequency passing through the
ferromagnetic transition of CeVSb$_3$ in various dc magnetic fields. The
peak amplitude monotonically decreases as field is increased.}
\label{cevsb3}
\end{figure}

\begin{figure}[htb]
\begin{center}
\includegraphics[width=9cm]{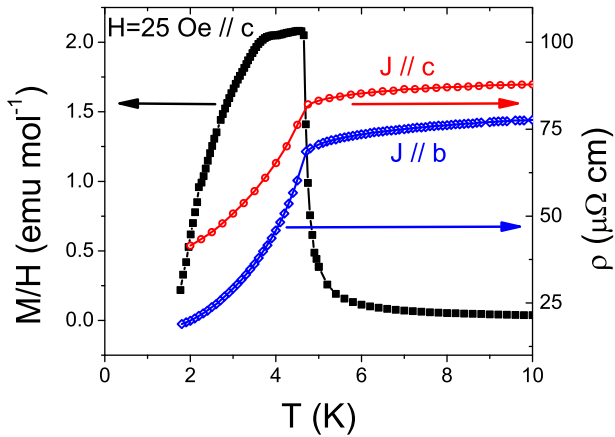}
\end{center}
\caption{(color online) Zero field cooled static $M/H$ (left axis) and
resistivity (right axis) along easy-axis (J//c) and in plane (J//b) for CeVSb%
$_3$ near $T_C$}
\label{cevsb3misc}
\end{figure}

Figure \ref{cevsb3misc} shows the low field, easy axis magnetization and the
resistivity measured in two in-plane directions: along the magnetic easy
axis and perpendicular to it. The perpendicular resistivity corresponds to a
skin depth ranging from 134 $\mu$m just above $T_C$ to 67 $\mu$m at 2 K for
our operating frequency. The dimensions perpendicular to the excitation
field of the sample used were approximately 400 x 100 $\mu$ m at the widest
point. This means the TDO data is relatively insensitive to changes in
resistivity because over the whole range the skin depth is greater than or
of the order of the size of the sample. This explains the relatively flat
response far from the ordering temperature.

\begin{figure}[htb]
\begin{center}
\includegraphics[width=9cm]{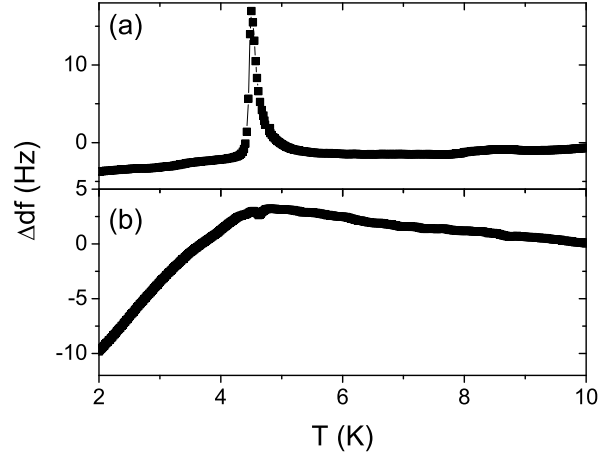}
\end{center}
\caption{Anisotropic zero field response of the TDO to CeVSb$_3$. Panel (a)
shows the response with the magnetic easy axis aligned with the excitation
field. Panel (b) shows the response with the magnetic easy axis
perpendicular to the excitation field. These measurements were performed on
a small piece of the original sample. The much smaller signal amplitude when
compared with Figure \protect\ref{cevsb3} is due to the smaller sample size.}
\label{cevsb3aniso}
\end{figure}

Figure \ref{cevsb3aniso} shows the zero field anisotropic response of the
TDO to a small piece of CeVSb$_3$ cut from the original sample. When the
magnetic easy axis is aligned with the excitation field ($H_{exc}$) there is
the sharp peak in susceptibility. However, when the $H_{exc}$ is
perpendicular to the easy axis no peak is present. From measurements on
CeAgSb$_2$ it is unlikely that the different responses in the different
orientations is due to demagnetization effects. The CeAgSb$_2$ data shows
similar responses separately for the FM axis aligned with and perpendicular
to the excitation field (see below). We do see anisotropic response, and
this will be studied further.

\begin{figure}[htb]
\begin{center}
\includegraphics[width=9cm]{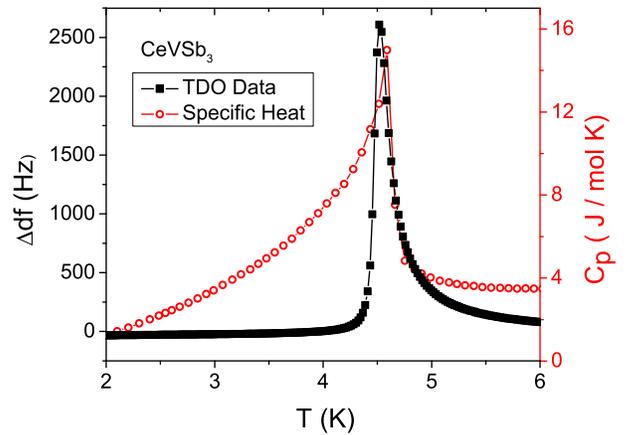}
\end{center}
\caption{(color online) Comparsion of zero field TDO data (left axis) with
specific heat (right axis)for $\mathrm{CeVSb_3}$. Solid squares are TDO data
and open circles are specific heat.}
\label{tdocp}
\end{figure}

Figure \ref{tdocp} shows the TDO response and specific heat measurement in
zero field in the vicinity of the phase transition. The TDO peak lies within
the specific heat peak. The latter peak is often taken as a demarcation of
the critical fluctuation region. This suggests that critical scaling
analysis may be applicable. Such analysis is presented in the discussion.

\subsubsection{$[Ce,Sm]AgSb_2$}

\begin{figure}[htb]
\begin{center}
\includegraphics[width=9cm]{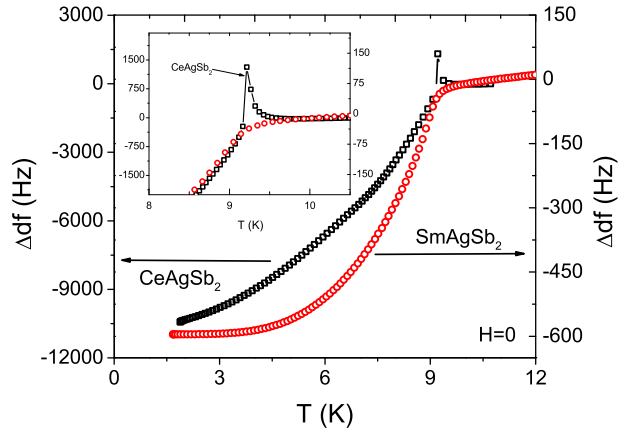}
\end{center}
\caption{(color online) Change in resonant frequency as temperature is
varied through the ferromagnetic transition of $\mathrm{CeAgSb_2}$ and the
antiferromagnetic transition of $\mathrm{SmAgSb_2}$. The sharp peak is
associated with $\mathrm{CeAgSb_2}$. Note both systems exhibit a decrease in
resistivity due to loss of spin disorder.}
\label{CeSmH0}
\end{figure}

\begin{figure}[htb]
\begin{center}
\includegraphics[width=9cm]{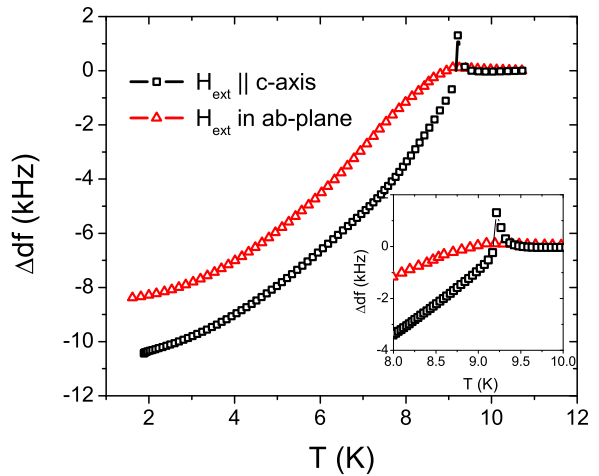}
\end{center}
\caption{(color online) Anisotropic response of the TDO to $\mathrm{CeAgSb_2}
$. Squares correspond to the c-axis aligned with the excitation field.
Circles correspond to the excitation field lying in plane. Some data points
are omitted for clarity. Inset: Detail of the phase transition.}
\label{ceagsbdir}
\end{figure}

Figure \ref{CeSmH0} shows the zero field frequency shift of the resonator in
response to $\mathrm{[Ce, Sm]AgSb_2}$. The cerium compound orders with a
ferromagnetic component along the c-axis at 9.8 K, whereas the samarium
compound orders antiferromagnetically at 10 K \cite{myers-1999}. Both
samples show a decrease in resistivity due to the loss of spin disorder
scattering, but the effect is much more pronounced in the cerium compound
(note different scales in figure). Since the samples have the same
dimensions, the most likely explanation of the differences in the
resistivity components is from the different changes in $\rho$ that occur
for each compound. CeAgSb$_2$ has a resistivity of approximately 90 $\mu
\Omega$ cm at 11 K which drops to about 1.2 $\mu \Omega$ cm at approximately
2 K. SmAgSb$_2$ has a resistivity of of about 1 $\mu \Omega$ cm at 11 K
which drops to about 0.5 $\mu \Omega$ cm at 2 K \cite{myers-1999}. Since we
directly probe changes in sample properties ($\chi$ and $\rho$) versus
temperature, the larger change in diamagnetic screening in the Ce compound
manifests itself as a larger shift in the resonant frequency of the circuit.
Further, the FM ordering of the cerium moments is accompanied by a sharp
peak in the resonator response. This peak is absent in the AFM SmAgSb$_2$.
Figure \ref{ceagsbdir} compares the zero field resonator responses of the
cerium compound in two orientations relative to $H_{exc}$. It is seen that
with the FM easy axis aligned with the excitation field, the susceptibility
peaks at the Curie temperature. However, the in-plane susceptibility shows
no such peak. As discussed above, the different demagnetization factors
along the magnetic easy axes of CeAgSb$_2$ (high value) and CeVSb$_3$ (low
value) do not affect the appearance of the peak at $T_C$. Therefore, it
seems unlikely that demagnetization is responsible for the disappearance of
the peak when the sample is oriented such that $H_{exc}$ is perpendicular to
the easy axis, rather it seems the TDO is sensitive to the magnetic
anisotropy of the sample.


\begin{figure}[htb]
\begin{center}
\includegraphics[width=9cm]{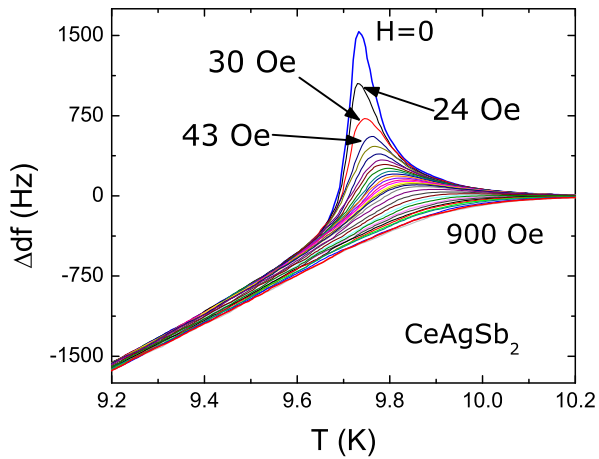}
\end{center}
\caption{(color online)Suppression of peak in susceptibility in the vicinity
of the ferromagnetic phase transition with applied magnetic field for $%
\mathrm{CeAgSb_2}$. This data, unlike others presented in this work, is
unshifted.}
\label{CeAgSb2peaks}
\end{figure}

Figure \ref{CeAgSb2peaks} shows magnetic susceptibility of the $\mathrm{%
CeAgSb_2}$ in applied dc magnetic field. Weak applied fields suppress the
peak dramatically and cause it to shift to higher temperatures. For fields
up to 900 Oe the resonator response far from the peak is independent of the
applied field. This is not surprising as the applied fields are too small
for magnetoresistance to play a significant role. Indeed, reported
magnetoresistance for this compound in the vicinity of the phase transition
shows less than 5\% change in $\rho$ for fields up to about 1 kOe \cite
{myers-1999} and our own measurements show no detectable change if $\rho$ in
this field range. Also, it is evident from the data that there is little
significant effect on the resonator response due to the applied field by
about 9.4 K for the region below $T_C$ and by about 10.2 K for the region
above $T_C$.

\subsubsection{$ZrZn_2$}

\begin{figure}[htb]
\begin{center}
\includegraphics[width=9cm]{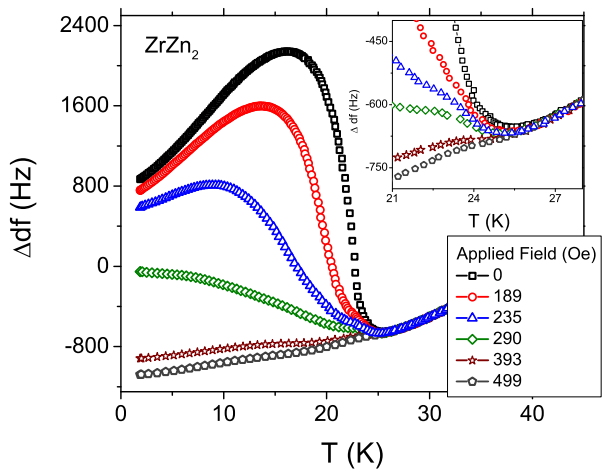}
\end{center}
\caption{(color online) Change in resonant frequency in response to the
itinerant ferromagnet ZrZn$_2$ in various magnetic fields. Inset: Detail of
response near $T_C$.}
\label{ZrZn2}
\end{figure}

ZrZn$_2$ is one of the prototypical weak itinerant ferromagnets ($T_C
\approx 28 \mathrm{K}$) \cite{yelland-2005}. Unlike CeVSb$_3$ and CeAgSb$_2$%
, ZrZn$_2$ does not manifest a sharp peak in the resonator frequency shift
at $T_C$. Rather, zero field data show a large increase in susceptibility
over a temperature range of approximately 5 K, a broad maximum, and a slower
decrease in susceptibility. At lower temperatures as the applied dc field
increases the temperature of the initial upturn in susceptibility does not
shift, however the broad maximum is shifted to lower temperatures and is
suppressed in amplitude. By 500 Oe the maximum virtually disappears (Fig. %
\ref{ZrZn2}). Also of note is the lack of a local maximum in the vicinity of
the phase transition.

Such extreme field dependence well below the transition indicates high
polarizability of the itinerant magnetic order and cannot be explained by
the direct contribution of the Brillouin variable $\mu H/k_B T$, where much
larger fields would be required and the peak would shift to higher
temperatures. Instead, it could be due to magnetic domains, which would then
be quite different in itinerant ferromagnets compared to the local-moment
systems and should be very weakly pinned, if at all. In this case, lower
temperatures would be required to create domains opposite to the applied
field, thus resulting in the maximum shift to lower temperatures.

\subsubsection{GdFe$_2$Zn$_{20}$}

\begin{figure}[htb]
\begin{center}
\includegraphics[width=9cm]{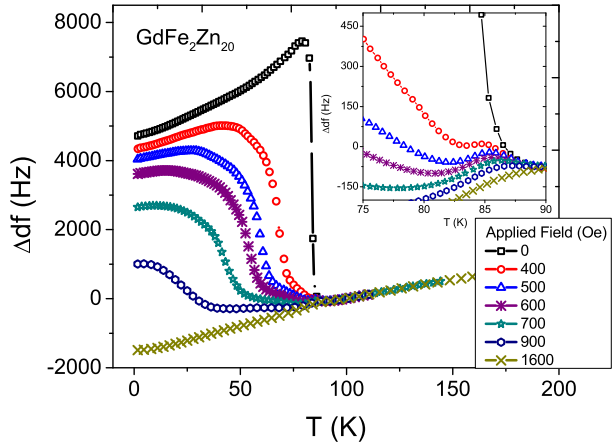}
\end{center}
\caption{(color online) Variation of resonant frequency with temperature for
$\mathrm{GdFe_2Zn_{20}}$. Inset: Detail of response near $T_C$. The small
peak is believed to be associated with the ordering of the localized
gadolinium moments.}
\label{GdFe2Zn20}
\end{figure}

GeFe$_2$Zn$_{20}$ is interesting for the present work as it contains both 4f
(local moment) and 3d (potentially itinerant) moment elements. This compound
develops a spontaneous magnetization below approximately 86 K, primarily
associated with the gadolinium ions. The conduction electron background is
highly polarizable (YFe$_2$Zn$_{20}$ being just below the Stoner limit) and
in the ordered state a 0.3 $\mu_B$ induced moment on the iron opposes the
gadolinium local moment \cite{jia-2006}. Figure \ref{GdFe2Zn20} shows the
measured TDO response due to this compound. The inset shows a detail of the
signal near the ordering temperature. The small local maximum close to $T_C$
exhibits behavior consistent with local moment ferromagnetic ordering, i.e.
suppression in amplitude and shifting to higher temperatures with applied
field. It is believed that this feature is associated with the ordering of
the gadolinium moments. The large feature on the main graph is reminiscent
of ZrZn$_2$, and therefore, believed to be associated with itinerant moments
or a band-like component associated with this relatively high ordering
temperature. AC susceptibility measurements on Pd-Mn alloys show strikingly
similar behavior (\cite{ho-1981} and \cite{ho-1981-2}). The data was
interpreted as domain motion and measurement saturation associated with
demagnetization effects. We believe the effect is from the separate
responses of the polarization of the conduction electron subsystem. We note
that a direct magneto-optical study of CeAgSb$_2$ (to be published
elsewhere) shows very soft domains at low temperatures, yet they do not
contribute to the low-temperature susceptibility as evident from Fig. \ref%
{CeAgSb2peaks}. This means, if itinerant domains are responsible for the
broad maximum in $\chi(T)$ in GdFe$_2$Zn$_{20}$ below $T_C$, they must be
very different from domains formed by local moments. A study of this
question is ongoing, including the construction of a new TDO to measure the
transitions in nickel, iron and other higher $T_C$ samples. Magneto-optical
techniques will also be used for direct observations of the magnetic
structure.

\section{Multiple Zero Field and Metamagnetic Transitions}

We now discuss two other compounds (Ce$_3$Al$_{11}$ and CeSb) where the TDO
was successfully used to measure changes in spin order as well as
metamagnetic phase transitions.

\subsection{Ce$_3$Al$_{11}$}

\begin{figure}[htb]
\begin{center}
\includegraphics[width=9cm]{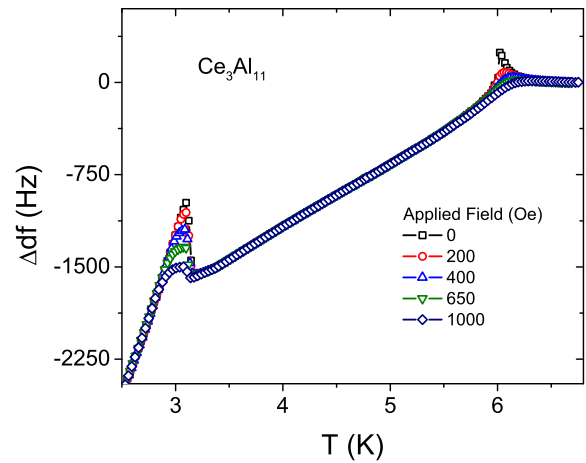}
\end{center}
\caption{(color online) Response of oscillator to $\mathrm{Ce_3Al_{11}}$ in
various magnetic fields through a ferromagnetic transition at 6 K and a
transition to an incommensurate state at 3 K. Inset: Close up of suppression
of ferromagnetic transition peak with applied field.}
\label{Ce3Al11}
\end{figure}

\begin{figure}[htb]
\begin{center}
\includegraphics[width=9cm]{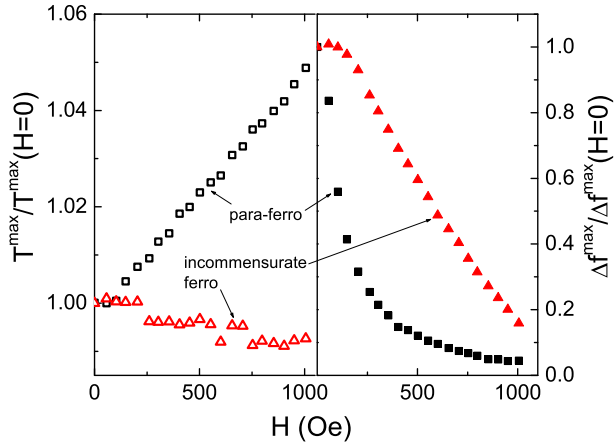}
\end{center}
\caption{(color online) Normalized temperature (open symbols, left axis) and
normalized amplitude (closed symbols, right axis) of susceptibility maxima
near the two phase transitions of Ce$_3$Al$_{11}$ relative to zero field
temperatures . Squares denote PM to FM transition while triangles denote FM
to incommensurate state transition.}
\label{ce3al11peaks}
\end{figure}

The metallic compound Ce$_3$Al$_{11}$ has a ferromagnetic transition at 6.2
K. Further cooling reveals another transition at 3.2 K to an incommensurate
spin state. In the FM state there are two distinct cerium sites carrying
different moments. The Ce$_I$ site carries a moment of 1.27 $\mu_B$ while
the Ce$_{II}$ site carries a moment of 0.24 $\mu_B$ as described by
Boucherle \emph{et al.} \cite{boucherle-1995}. TDO measurements clearly
detect both transitions as shown in Fig. \ref{Ce3Al11}. The para- to ferro-
transition is characterized by the familiar sharp, asymmetric peak that
suppresses in amplitude and shifts to higher temperatures with increasing
applied field. The transition to the incommensurate state is marked by a
rapid increase in the susceptibility followed by a decay as temperature
continues to drop (Fig. \ref{ce3al11detail}). Application of magnetic field
does not induce a significant temperature shift in this lower transition
(Fig. \ref{ce3al11peaks}), however it does trend slightly to lower
temperatures, and the amplitude of the susceptibility increase is
suppressed. It should be noted that far from the transitions, the measured
signal is insensitive to applied fields, which means it is likely that the
differences in response from one field to another we see are only due to the
magnetic and not the skin effect component of $\chi$.

The resistivity of Ce$_3$Al$_{11}$ varies from 20 $\mu \Omega$ cm at 7 K to
5 $\mu \Omega$ cm at 2.5 K. Published data shows that the resistivity is
nearly constant from 7 K down to the FM transition where it exhibits a steep
drop. Further, there is a change in slope in the resistivity versus
temperature at the transition to the incommensurate state \cite{ebihara-2004}.
This corresponds to a skin depth at our operating frequency variation of
approximately 68 $\mu$m at 7 K to 33 $\mu$m at 2.5 K. The sample used in
this experiment had dimensions perpendicular to the excitation field
approximately six times larger than the largest skin depth. Therefore, the
TDO should be sensitive to the changes in $\rho$ over temperature range of
interest. This is consistent with the results presented here.

\subsection{$CeSb$}

\begin{figure}[htb]
\begin{center}
\includegraphics[width=9cm]{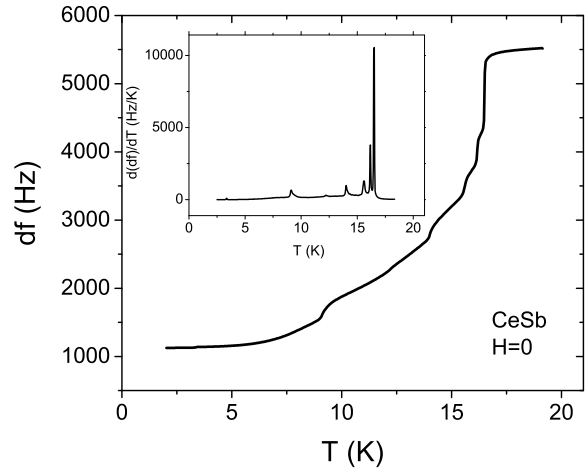}
\end{center}
\caption{Zero field response due to various magnetic transitions in CeSb.
Inset: Derivative plot of main graph. The small peak at $\sim$ 1.8 K is due
to the suppressed superconducting transition of a small amount of tin flux
residue from the crystal growth process.}
\label{CeSb}
\end{figure}

\begin{figure}[htb]
\begin{center}
\includegraphics[width=9cm]{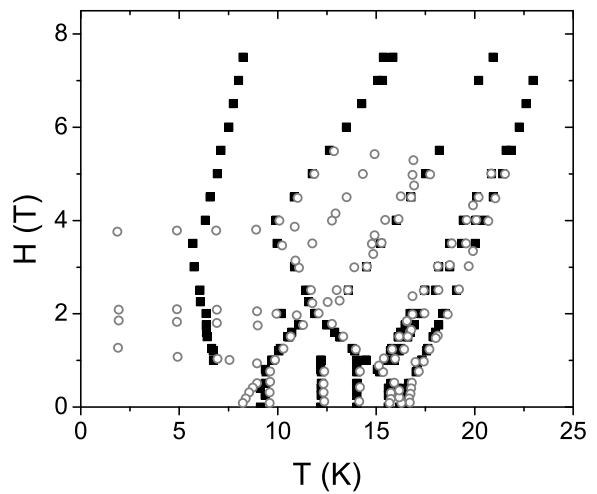}
\end{center}
\caption{$H-T$ phase boundary plot for CeSb obtained from TDO response.
Solid squares are this work, open circles are from Wiener and Canfield
\protect\cite{wiener-2000}.}
\label{CeSbpd}
\end{figure}

Given that the TDO seems to be sensitive to multiple transitions, we decided
to test this sensitivity with the binary compound CeSb, which has an
extremely rich $H-T$ phase diagram with many different ordered phases \cite{wiener-2000}.
Each phase transition has an associated change in the long
range magnetic order which is evident (to varying degrees) in thermodynamic
and transport measurements. The changes in $\chi$ are readily seen in the
raw data from the TDO (Fig. \ref{CeSb}). It should be noted that TDO data
alone will not, in general, give the actual ordering of any particular
phase. Its use lies in its ability to detect phase changes with extreme
sensetivity. Temperature sweeps from 1.5 - 20 K were performed in 25
different applied fields ranging from 0-7.5 T. Temperatures of the peaks of
the derivatives of the frequency shift were recorded for each field. The
generated \emph{H-T} phase boundary plot (Fig. \ref{CeSbpd}) matches that
reported in reference \onlinecite{wiener-2000} very well. Noticeably absent
are the field induced transitions near 1, 2, and 4 T previously reported.
Given that in the present work no field sweeps at constant temperature were
taken, this is not surprising. It is likely that these boundaries would be
seen by a TDO under the proper experimental procedures. Two other
differences between this work and previous studies is here we measure a
boundary that lies between 5 and 7 K for fields between 1 and 7.5 T, and we
do not see the transition between two different ferro/para mixed states
running along a line from about 12 K, 3 T to about 15 K, 5 T. The latter
difference may be due to a perpendicular alignment of the ferromagnetic
moment with respect to the excitation field.

\section{Scaling analysis of magnetic susceptibility}

Finally, to demonstrate quantitative analysis capabilities, we present
initial scaling analysis of our CeVSb$_3$ TDO data.

\begin{figure}[htb]
\begin{center}
\includegraphics[width=9cm]{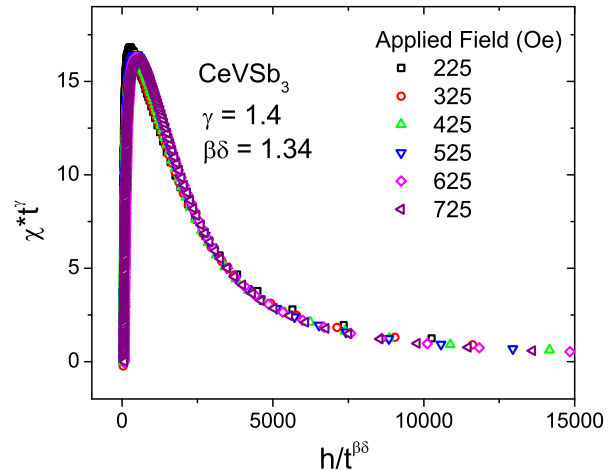}
\end{center}
\caption{(color online) Plot of scaled susceptibility vs. scaled field for
CeVSb$_3$. The temperature range shown is from 12 K to 4.6 K left to right.}
\label{scaling}
\end{figure}

In the vicinity of a phase transition the correlation length over which
local order exists diverges. This leads to the variation of thermodynamic
quantities (\emph{e.g.} specific heat or magnetic susceptibility) being
dominated by power law behavior in either temperature or field. The field
and temperature region over which this happens is the critical region, and
the exponents are critical exponents or critical indices. Relevant for
magnetic susceptibility measurements are the exponents $\beta$, $\delta$,
and $\gamma$ defined by
\begin{align}
m \sim \ t^{\ \beta} \\
m \sim h^{1/\delta} \\
\chi \sim t^{-\gamma}
\end{align}
Here $t$ is the reduced temperature and $h$ is the reduced field.
\begin{align}
t=1-{\frac{T}{{T_C}}} \\
h={\frac{\mu H }{{k_B T_C}}}
\end{align}
In both of the above expressions $T_C$ is defined as the zero field
temperature of the phase transition and $m$ is the magnetization per unit
volume. The magnetic moment per ion is $\mu$ and $k_B$ is the Boltzmann
constant. It is straightforward to show that
\begin{equation}
\chi \sim h^{(1/\delta)-1}
\end{equation}
Analysis of critical phenomena rely on the assumption that the power law
behavior completely dominates the system as the transition temperature is
approached from below or above. The static scaling hypothesis requires that
the free energy is a generalized homogeneous function in two variables. This
leads to the conclusion that only two of the critical-point exponents need
to be determined. All others are derivable from various scaling relations.
Further, under the static scaling hypothesis the exponents below $T_C$ are
equal to those above. Under the scaling hypothesis, the exponents $\beta$, $%
\delta$, and $\gamma$ are related by the following equation
\begin{equation}
\gamma = \beta(\delta-1),
\end{equation}
from which we also find
\begin{equation}
\beta \delta = \beta + \gamma.
\end{equation}
The above relations are based on the magnetic equation of state
\begin{equation}
m(h,t)=t^\beta F(h/t^{(\beta \delta)}).
\end{equation}
Differentiating with respect to $h$ gives the susceptibility equation
\begin{equation}
\chi(h, t)=t^{-\gamma} \dot{F}(h/t^{(\beta \delta)}).
\end{equation}
Measurements of susceptibility in various fields near $T_C$ should fall on a
common curve if plotted as $t^\gamma \chi$ versus $h/t^{\beta \delta}$ (Fig. %
\ref{scaling}.) The value of $\gamma$ can be estimated by fitting a line to
a log-log plot of the TDO response vs. reduced temperature near $T_C$. Even
though our system has an unknown offset in $\chi$, the response should be
dominated by the diverging component related to the phase transition. The
resulting value of $\gamma$ from our data is 1.4. The product $\beta \delta$
can be treated as an adjustable parameter. We find $\beta \delta = 1.34$
results in a good collapse of the data. While the value of $\gamma$ is
consistent with a 3D Heisenberg model, the product $\beta \delta$ is closer
to the prediction of a 3D Ising model.

\section{Conclusions}

Sensitive tunnel-diode oscillator techniques were used to study magnetic
susceptibility in several intermetallic magnetic compounds. It is
demonstrated that the TDO has strong potential as an instrument with which
to study magnetic phase transitions. Ferromagnetic transitions are of
particular interest as the behavior of the susceptibility peak in the
vicinity of $T_C$ is largely inaccessible by other means. The anisotropic
response seems to allow one to measure the ferromagnetic easy axis of a
compound in the zero field limit near $T_C$. The possibility to detect the
difference in the dynamic response between weak itinerant and local moment
magnets is very encouraging, because it is often difficult to determine the
mechanism of magnetism in the ordered state. Quantitative analysis of the
response was demonstrated for critical scaling near $T_C$.

\begin{acknowledgments}
Work at the Ames Laboratory was supported by the Department of Energy -
Basic Energy Sciences under Contract No. DE-AC02-07CH11358. R. P.
acknowledges support from the NSF grant number DMR-06-03841 and the Alfred
P. Sloan Foundation.
\end{acknowledgments}

\bibliographystyle{plain}
\bibliography{TDR}

\end{document}